%% file: ontoPhysics.tex
\documentclass[numbers=noendperiod,chapterprefix,sopenany,fontsize=11,oneside,bibliography=totoc,parskip=half-]{scrartcl}
\usepackage{epsf, pstricks, graphics, graphicx} 
\usepackage[a4paper]{geometry}
\usepackage{multirow} 
\usepackage{tikz}
\usepackage{dot2texi}
\usepackage{svg}
\usepackage{physics}
\usepackage{endnotes} 
\usepackage[edges]{forest}
\usepackage[colorlinks=true, urlcolor=blue, linkcolor=blue]{hyperref} 
\hypersetup{colorlinks = true, urlcolor = blue}
\usepackage{xfrac} 
\usepackage[backend=biber, citestyle=numeric, bibstyle=numeric,sortcites]{biblatex}
\usepackage{float} 
\usepackage{pbox} 
\usepackage{hyperref} 
\usepackage{threeparttable}
\usepackage{rotating} 
\usepackage{setspace} 
\usepackage{rotating}
\usepackage{multicol} 
\usepackage{tablefootnote}
\usepackage{cleveref}
\usepackage[T1]{fontenc} 
\usepackage[utf8]{inputenc}
\usepackage{amsmath}
\usepackage{mathtools}
\usepackage{amssymb}
\usepackage{xfrac}
\usepackage{mathcomp}
\usepackage{lscape}
\usepackage{imakeidx}
\usepackage{tabularx}
\usepackage{multicol}
\usepackage{framed}
\usepackage[german,english,polutonikogreek]{babel}
\newcommand{\x}[1]{\textit{#1}}
\newcommand{\y}[1]{\textbf{#1}}

\begin{document} 
\selectlanguage{english}
\title{A Husserlian ontology of the science of physics}
\author{Jobst Landgrebe \and Barry Smith}
\date{\today}
\publishers{University at Buffalo\\
jobstlan@buffalo.edu, phismith@buffalo.edu}
\maketitle

\begin{abstract}
Since 1939, when Wigner published his proposal to define particles via symmetries expressed as algebraic groups, we have seen a long stream of attempts to formulate an ontology of matter based on mathematics. It has  become apparent that such attempts must fail, and more particularly that we cannot derive an ontology of matter from the Standard Model on the basis of quantum field theory. We briefly recapitulate the reasons for this and demonstrate how the philosophy of physics has become scholasticised through a series of vain attempts to obtain a coherent ontology of matter. Here we propose an alternative approach in the form of an ontology of the science of physics. It draws on a framework developed by Husserl which admits not only real material entities but also the idealities which physicists make use of in their thinking. We specify the upper level of this ontology, which addresses the way in which physicists use mathematics when conducting their enquiries. This ontology provides a new perspective concerning the dilemmas of modern physics, including the measurement problem. 
\end{abstract}

\input{intro}
\input{results}

\input{discussion}
\input{appendix}

\printbibliography

\end{document}

%% file: intro.tex

\section{Introduction}

A formal ontology provides the formal categories under which all entities fall, together with the associated formal definitions and axioms. Ontology in this sense is a part of metaphysics; it is the study of the most fundamental properties of being. Formal ontologies are maximally general and are associated with material ontologies at lower levels. A \x{material ontology of physics} attempts to provide the resources to answer questions about the nature of matter, space and time, and also to formulate the general principles that govern the interactions of matter to yield the phenomena we observe in nature. 

Our starting point in this exercise is the work of Husserl, who was the pioneer of modern ontological thinking. Thus it was he who introduced the term `formal ontology' already in his third Logical Investigation \cite{simons:1992,smith:1986,smith:2010,smith:2022}.

The later Husserl classifies all entities into three types. There are real entities at the level of both universals and their instances, which are the particulars we find in the world we live in. These instances include our bodies, the processes in which our bodies participate, and associated qualities and functions for example of their constituent cells. All real entities at the level of instances are at any given time located in some region of space.

In addition, however, there are non-real (Husserl calls them \x{irreal}) entities, which have no individuation via a location in spacetime. These fall into  two groups, which Husserl called \x{free} and \x{bound} idealities \cite[p.~97]{erhard:2014}. We will utilise these fundamental entity types in what follows.

\subsection{Universals}

Universals, for Husserl, are real entities, by which he means that they have instances which are `individualised through their position in spacetime' \parencite[p.~116]{husserl:1966}. Examples of such instances are this tree, that lion, a face, a role (for example the role of being a sibling), or a quality (such as the redness of this apple), and so forth. Such individuals belong to the same universals as other, like individuals because they share essential features. For example, deciduous trees are trees that lose their foliage, while coniferous trees are evergreen. We see from this example that universals are defined using the Aristotelian genus-species approach to definition. This defines each universal in terms of  the more general universal which is its parent (in this case: \x{tree}) by adding thereto another property (a specific difference). According to Husserl, knowledge of universals is obtained by abstracting from individuals \parencite[Inv.~6, \S 60]{husserl:2000}, while for Aristotle,  such knowledge exists independently of abstraction. This means that from the Husserlian perspective, universals depend on their instances (individuals), since they are given to us by abstraction from them.

\subsection{Free idealities}

Let us first of all review Husserl's view of the real and the irreal. He says:

\begin{quote}
We call ‘real’ [$\dots$] everything that is essentially individuated through its spatiotemporal location, ‘irreal’ every determination whose spatiotemporal appearance has a foundation in some specific real entity but which can appear as identical – and not merely similar – in different realities \parencite[p.319]{husserl:1999}.
\end{quote}

The irreal has no instances, and no individuation via a position in spacetime. An irreal entity is everywhere and nowhere. Husserl gives two examples that show us what he means by `has a foundation'. The first example is a cultural entity, Goethe's \x{The Sorrows of Young Werther}. Though it exists in many printed versions the work of art itself is an ideality. The foundation of this ideality are the real printed copies that allow us to read the novel. As an ideality the novel has an existence of its own, which arises in the mind of the reader during the reading process.\footnote{See \textcite{ingarden:1973} on what this process involves.} Husserl's second example are propositions from geometry, which can be uttered or written down but are nevertheless independent of any given representations. Thus we can say that what Husserl means when he talks of the `the foundation' of idealities `in a real entity' is the need for real-world representations (utterances, texts, images, or other symbols), through which individuals can become acquainted with them.

Free idealities are irreal entities which come close in the history of philosophy to what Plato called `ideas', which have of course given rise to endless debates. For Husserl, there are free idealities, such as the circle or the triangle, which we find in mathematics. These do not require any bearer for their existence; they are in themselves self-contained, self-sufficient abstract entities. Further examples are $\pi$, the Euler number $e$, or the imaginary number $i^2=-1$. All of these are free idealities, which means that they are not bound to -- thus not dependent on -- our sensations or thoughts, and therefore also not to a specific culture. The ideality of a geometrical object is not derived from our assertions about the object, but is a feature of the object itself. Each such ideality is unique, but it can be rediscovered by the human mind \x{ad infinitum}: \begin{quote} The Pythagorean theorem, indeed all of geometry, exists only once, no matter how often or even in what language it may be expressed. $\dots$ and within each language it is again the same, no matter how many times it has been sensibly uttered, from the original expression and writing-down to the innumerable oral utterances. \cite[\S 65]{husserl:1975}. \end{quote}

Next to mathematical entities, Husserl (like his colleague Scheler \cite{scheler:1973}) accepts also values and norms as free idealities.

\subsection{Bound idealities}

For Husserl \x{bound idealities} are defined by the fact that to be what they are supposed to be they require a real or phenomenal bearer. The letter-form A, for example, is bound to sensible appearances.

He explains the difference between free and bound idealities as follows \parencite[loc. cit.]{husserl:1988}:

\begin{quote} It appears that even cultural systems are not always completely free idealities, and this reveals the difference between free idealities (such as logicomathematical systems and pure essential structures of every kind) and bound idealities, which in a sense carry reality with them and hence belong to the real world. 
[$\dots$] Free idealities are bound to no territory, or rather, they have their territory in the totality of the universe and in every possible universe. In what concerns their possible reactivation, they are omnispatial and omnitemporal.\end{quote}

Bound idealities, in contrast, are bound to Earth, to Mars, and so forth. They are general in application but have no instances in the way universals have. Instead they are tied to individual entities -- such as the letter `A' or the US Constitution --  via approximation, and also via convention. For example, the units of measurement used to quantify mass, such as the unit gram, are a product of convention, while mass itself is a universal with instantiations in spacetime.

Both free and bound idealities, in sum, are distinguished from universals by the fact that the latter have (or could have) instances in reality. Among the examples of bound idealities provided in Husserl's writings are \x{phonemes} and \x{letters.} Their identity is ideal, but they require inscription or articulation to be perceived, which is why they are bound to reality. 
The English word `tree' exists as a universal through its many instances in spoken or written form. The meaning of a word as an ideality (an irreal entity) is bound to its physical or symbolic presentation (for example ink marks or sound waves) realized wherever the word is used. Husserl also uses social norms as examples of bound idealities, pointing out for example that `The constitution (\x{Verfassung}) has an ideality, insofar as it is a categorical objectivity it is repeatable at different times.' \cite[p.~320]{husserl:1999}.

\subsection{The BFO upper ontology}

There are many domain ontologies which supply the universals under which entities such as a finger, a tree, or a house can be classified at different levels of generality. These domain ontologies require an upper ontology whose terms form the starting point of Aristotelian definitions. One such upper ontology is BFO (for: Basic Formal Ontology \cite{bfo:2020}), a realist ontology focused primarily on real entities, for example as studied by the natural sciences. The totality of real entities are divided, for BFO, into continuant and occurrent entities. BFO comprises a taxonomy of universals that have real particulars as their instances. These universals together form a pattern -- of objects, processes, qualities, regions of time and space, and so forth, connected by relations of different sorts -- and it is our contention that BFO comes close to classifying -- albeit at a very high level of generality --  the entities and relations that dominate our understanding of the common-sense world.\footnote{See: https://basic-formal-ontology.org/users.html/.}

However, BFO in its current version has no place for free and bound idealities, for here the universal-instance pattern does not hold. Examples of relevance to physics are, in addition to $\pi$, a Riemannian manifold or a Hilbert space.  From this it follows that BFO's claim to be a top-level ontology\footnote{ISO/IEC 21838-1:2021 Information technology -- Top-level ontology (TLO), Part 1: Requirements; Part 2: Basic Formal Ontology.} needs careful scrutiny. Either BFO itself needs to be extended to encompass mathematics and the sciences based thereon. Or BFO needs to be viewed as a sub-ontology of a more all-encompassing ontology ecosystem comprehending entities in these domains. Or, finally, good reasons need to be provided why mathematics and especially quantum physics do not merit a treatment in ontological terms at all, because there are strictly speaking no entities in these domains. 


Attempts at creating ontologies in the realm of physics are thus far unsatisfactory. The aim of this paper is to propose an ontology of the science of physics and its objects that is complete, consistent and compatible with common sense as far as physics allows it.

\subsection{Outline of this contribution}

All ontologies of physics have in common that they inherit the interpretation problems of quantum mechanics (as also of thermodynamics and the general theory of relativity), as well as many further problems pertaining to the theories of matter which are associated with the existing variants of quantum physics \cite[ch.~7]{landgrebesmith:2025}. We can, to be sure, calculate very exactly with the equations of quantum physics in such a way as to obtain probabilities. The problem is that we do not understand  -- in Feynman's sense of the word `understand' \cite{feynman:1960} -- what it is that we observe when we perform the associated experiments. Over the last 100 years, we have derived mathematical models from these experiments, and these models allow us to predict very accurately the results of further experiments. But these models do not allow us to derive an ontology of nature  -- that is, they do not allow an accounting of the entities that exist in the physical world that they are models of. Rather we shall argue that the mathematical models of modern physics, if they are to be adequate to physics as a science, necessitate an ontology like the one proposed in this communication.

In what follows we first summarise the most important attempts to propose ontologies of physics over the last decades. Many of these conflate the physical objects with the models of such objects created by physicists, which is metaphysically incoherent. A picture of the sun is not the sun, and an Einsteinian manifold is not the real space and time that we perceive. 

We will then, using examples from quantum physics, describe why a common-sense oriented ontology like BFO  cannot as it stands be used as a formal ontology covering the physics domain. We then propose a new approach to the ontology of physics, which proceeds by providing an account of the \x{structure of the science of physics}, consisting of three branches:

\begin{enumerate}
\item systems, system elements, and system relations,
\item magnitudes used to measure properties of system elements and relations,
\item models, which are constituents of theories.
\end{enumerate}



%% file: results.tex
\section{Context and motivation}

To understand why existing approaches to the ontology of physics fail, we first summarise important examples of such ontologies and then look more closely at examples from quantum mechanics.

The aim of existing ontologies of physics is to describe (1) the entities (such as planets, atoms, or bosons) to which the scientific inquiries of the physicist are addressed and (2) the models and theories resulting from these inquiries.
These are, of course, very different types of entities. The first is always a system\footnote{A system is a totality of dynamically interrelated physical elements participating in a process \cite[Glossary]{landgrebesmith:2025}. See \ref{sysE} below.} of entities selected by the physicist for enquiry, while the second, a model (or theory such as the general theory of relativity), is the ultimate result of applying the practical and theoretical methods established in the discipline. The practical methods include apparatus construction on the one hand and experimental measurements or observations on the other, where the latter yield data about the system to which the model applies. Theories in quantum mechanics apply exclusively to man-made experimental settings or devices engineered on the basis of previous quantum theoretical models. Such devices can be seen as experiments in practical usage -- see section \ref{disc}.

There is a multitude of ontologies of physics. In many of these, nature (1) and models/theories (2) are conflated into a single ontology, as if the physical reality and the results of practicing physics were identical. Their creators are then often not aware of the difference between physical \x{reality} and \x{physical} theory. For example, many try to formulate an ontology of space and time based on the results of the general theory of relativity by utilising Einsteinian time-space manifolds as if the latter were elements of real systems -- which they are not. They are mathematical models. This misunderstanding drives a huge branch of metaphysical theorising called `four-dimensionalism' \cite{sider:2001}. Its proponents seriously believe that a mathematical formalism\footnote{The formalism that is required to formulate Einstein's field equations.} shows us how the phenomena \x{are}, though the purpose of the formalism is to \x{model} physical phenomena at high speeds or high masses, not to \x{be} them. A blatant example of this misunderstanding can be found in \textcite{huggett:2013}, whose authors use speculative mathematics of quantum gravitation theory to drive an ontology of what they call `emergent spacetime'.

This metaphysical confounding of models of reality with reality itself prevails also in the ontologies of physics concerned with the nature of matter. Since the world is made of matter, most ontologies of physics focus on the theory of matter, starting out from the Standard Model of particle physics, and attempting to derive an ontology therefrom.
An overview of the existing ontologies of physics is given in Appendix (A), we focus here on one characteristic contribution to this literature.


In the volume edited by \textcite{kuhlmann:2002}, Peter Simons gives a comprehensive overview of the leading potential ontological frameworks which might be used to represent QFT, all of which are ontologies of one or other \x{theory} and of the theories' models that equate the theory and its models \x{with nature} itself. For example, Simons asks the following questions:

\begin{quote}
Are waves or fields fundamental and particles aspects thereof \footnote{For `aspects', Simons uses `moments'.} or the other way around? Or are both dependent on some third thing? Is spacetime fundamental and matter-energy a moment or the other way round? Or are both dependent on some third thing? [$\dots$] There is not one message coming from physics as to its ontology, but several at once. This is another reason for not wanting to craft one's ontology solely on the basis of this or that current physical theory. \cite[pp.~46f.]{simons:2002} \end{quote}

Here, Simons asks ontological questions about the reality of mathematical entities. But he does not see that the entities he is writing about are idealities -- which we described in the foregoing. Therefore the entire perspective presented by Simons is beside the point, since all QFT entities, including the particles he mentions, are not real entities, but rather bound or free idealities. In other words, all the potential ontologies of QFT presented by Simons struggle with an inadequate account of the nature of the models: they confound models (which are made up of idealities) with reality. But while it is clear that even though all QFT models are about an aspect of reality, the reality of measurement results, we shall see that we can only model the reality specific to the quantum phenomena themselves using idealities. The QFT models themselves consist of idealities although they are related to measurement results.

The deep reason for this confusion is that contemporary Anglo-Saxon philosophy ignores the existence of idealities. This is also evident from Simons' view that there is `not one message coming from physics'. For there is indeed only one such message, in spite of the different formulations of QFT, as we shall see in sections \ref{einst} and \ref{onePT}. However to understand this message, one has to see all the entities of modern physics, namely systems, magnitudes, and models for what they are, namely idealities.


\subsection{Einstein's dilemma} \label{einst}

To introduce our ontological framework, we give two examples from quantum physics. We begin with Einstein's famous thought experiment to show the incompatibility between the two postulates of completeness and locality. The first postulate is that the quantum mechanical description of reality is complete, the second that physics describes only processes with local interactions. We follow \textcite{norsen:2005} in his description of the thought experiment and provide an interpretation based on our ontology in section \ref{dilemma}.

In 1927, Einstein described a particle moving towards a diaphragm with a narrow aperture and a hemispherical detection screen behind it \parencite[~166 and Fig.~2]{norsen:2005}. If a single electron is sent through the aperture, it will diffract and move as a spherical wave described by the Schr{\"o}dinger equation towards the screen with a spread-out probability to reach the screen somewhere. But it will nevertheless be detected as a point on the screen. The equation describing its movement in time is:

\begin{equation}\label{eeq}
i \hbar \frac{\partial }{\partial t} \psi(x) = \hat H \psi(x),
\end{equation}

where the usual definitions of the variables apply \parencite[vol.~3, ch.~16-5]{feynman:2010}.

The equation describes a wave without giving information about an individual particle (in our example, an electron). Though from a classical perspective the particle follows a trajectory in spacetime, the model represents an ensemble average over a large number of individual trajectories (ensemble view, Norsen, loc. cit.). But upon detection, the electron is measured as a single particle. This second (measurement) view of the theory also `has the pretension to be a complete theory of individual processes' \cite{ballentine:1972}, where $|\psi|^2$ gives the probability that a single particle has a certain location (just before measurement). This is called the collapse of the wave function.

The ensemble view is incomplete because it fails to model the trajectory of each individual particle (as opposed to the Hamiltonian trajectory of classical physics) between the slit and the detection screen, and the second view ($|\psi|^2$-view) leaves us unable to assign a trajectory to each individual particle since it only gives a probability density. Now Einstein points out that the particle is always found at a certain point on the detector. But if we hold that `the initially spherical wave accurately and completely captures the physical reality of the propagating particle, this collapse evidently involves a kind of action-at-a-distance' \parencite[p.~167]{norsen:2005}: This is because as long as it is not detected, the particle has an almost spherically distributed probability to be detected anywhere on the screen, but when it is detected `a peculiar action-at-a-distance must be assumed to take place which prevents the continuously distributed wave in space from producing an effect at \x{two} places on the screen' \parencite[p.~116]{jammer:1974}. Given all of this, Einstein's thought experiment shows that `any possible interpretation in which the wave function was regarded as a complete description of physical reality would have to entertain the unacceptable (reality-violating) action-at-a-distance associated with the collapse of the wave function on measurement.' (Norsen, loc. cit.) Since Einstein was not willing to accept non-locality, he thought that the wave function is incomplete. As we shall see, he was right (sect.\ref{dilemma}).

It can be argued that Einstein's dilemma merely arises from a thought experiment. But later on Heisenberg proposed an experiment that allows one to actually measure the dilemma. If we imagine a wave packet of photons sent through a semi-reflecting mirror, it should be possible to split the package in two parts, one that goes through the mirror and one that does not. If this experiment is done with one photon, its quantum mechanical description before the measurement is:

\begin{equation}\label{mirror}
  \ket{\psi}=\frac{1}{\sqrt{2}}(\ket{\psi_1}+\ket{\psi_2}),
\end{equation}

where the subscripts $1~\mathrm{and}~2$ indicate the side of the mirror where the photon will be detected after arriving at the mirror. Such an experiment was indeed performed. It was found that a single photon is detected on one side upon every measurement; it never splits \parencite[p~172f.]{norsen:2005}. Because equation (\ref{mirror}) cannot attribute a position to the particle, it is incomplete. \x{However, by assuming incompleteness, we can avoid non-locality.} On the other hand, if we uphold the thesis that the model is complete, we have to deny the locality principle and regard `the collapse of the wave function on measurement as describing a real physical change in the state of the particle.' (Norsen, loc. cit).

Thus also this Heisenberg-inspired experimental validation of Einstein's dilemma shows us that we must either assume locality or abandon the completeness of the wave function as a description of reality (Einstein). What does this mean ontologically? We shall see this in section \ref{dilemma}.

\subsection{One photon from two sources} \label{onePT}

We now turn to another example of the ontological problems following from quantum physics.

Classical physics, quantum mechanics, quantum electrodynamics, quantum field theory, and solid-state physics all  conceive particles differently \parencite[ch.~6]{falkenburg:2007}. The particles of classical physics have characteristics which are in perfect agreement with our natural view of the world and were developed using a combination of common sense and mathematics. They have the meronymic (parthood) property of forming systems and the property of interacting with each other, which allows us to model motion. In classical physics, particles are carriers of mass and charge, are independent of each other, have pointlike interactions, are subject to conservation laws, are completely causally determined by the laws of mechanics, move on Langrangian trajectories in phase space, are spatio-temporally individuated, and can form bound systems (such as molecules).

But the particles of quantum physics lack many of these properties \cite{falkenburg:2007}. Particles of quantum mechanics are localisable only by applying measurement operators to the wave function. Without these operators, which are mathematical models of the corresponding experimental particle detectors, nothing happens according to quantum mechanics, other than the motion of the quantum wave as described by Schr{\"o}dinger's linear differential wave equation. Furthermore, magnitudes of quantum particles can only be determined according to the Heisenberg uncertainty relation. The particles behave like waves displaying the phenomena of superposition and interference. They cannot be individuated in the commonly understood sense, though there are of course multiple experimental settings in which single particles can be detected (for example one particle before and after a slit) if the individual measurement of the magnitude is not relevant.

With regard to magnitudes such as spin, position\footnote{On Bohm's theory of position, see p.~\pageref{bohm} below.}, momentum, or energy, the equations of quantum mechanics are merely probabilistic and apply to ensembles of events detected when measuring these magnitudes. There are -- strictly speaking -- no classical individual particles in quantum mechanics \x{when it comes to the experiments beyond mere detection or involving quantum phenomena such as entanglement, world-lines, or wave-particle-dualism.} The particle tracks which we measure in particle detection devices are calculated using classical physics (by means of the Lorentz force), but the energy loss of the particles along the track is a quantum phenomenon which cannot be described adequately using the classical Lagrangian world line.

What holds classical and modern physics together is the correspondence between the quantum and the classical world formulated by Bohr and extended by Heisenberg, which postulates a detailed analogy between the quantum and the classical theory. But this correspondence can be maintained in certain situations only, for example for the case of Rutherford scattering with slow-moving charges and in the absence of spin or quantum interactions. But in general it breaks down \parencite[Ch. ~5.4.2]{falkenburg:2007}: `There are many quantum phenomena without classical correspondence.' 

To exemplify the situation of total break-down of the meronymic and causal particle concept in quantum mechanics, consider the famous Pfleegor-Mandel experiment from 1967. When Pfleegor and Mandel prepared photons using two separated lasers to interfere in coherent phase, a wave interference pattern resulted as expected from the double slit experiment. But when the lasers were then attenuated so much that in the field of the experiment there remained only one photon, a weak interference pattern was still detected although the number of photons was only 1. Because of the interference this photon\footnote{Strictly speaking we must say that the `light intensities are so low that, with high probability, one photon is absorbed before the next one is emitted by one or the other source' \parencite[sect.~7.3.1]{falkenburg:2007}.} must have been created from the superposition of two field modes with occupation numbers 0 and 1 with zero photons from the first and one photon from the second laser and vice versa \parencite[sect.~7.3.1]{falkenburg:2007}

The superposition equation for the two wave functions is:
\begin{equation}\label{e:two} 
\Psi_1 \otimes \Psi_2 = \frac{1}{\sqrt{2}}(\ket{1_1}\ket{0_2} + \ket{0_1}\ket{1_2}),
\end{equation}

where $\ket{1}$ means state laser on and $\ket{0}$ means state laser off, while the subscripts designate the two lasers.

Because of the interference pattern, the photon must have been created by both lasers, in other words, both lasers emitted a quantum wave, though only one photon was detected. This one photon is made of contributions from both lasers. The detected photon has no path in spacetime and no causal provenance. It also is not clear how it is formed so that it has no meronymic character either. As Falkenburg points out, \x{the correspondence with classical physics completely breaks down, the photon cannot be understood as a particle in the classical sense, though it can be detected as such} (loc. cit.)

How are we to interpret this in a framework of common sense based on the way we humans understand the reality that we live in? None of the ontologies of physics described above is of any help there. To remedy this situation, starting from this example, we proceed to lay out an ontology of physics that enables an adequate interpretation of these phenomena.

\section{The realist ontology of physics}\label{realO}

The realist ontology of the science of physics we propose is structured in correspondence with the way knowledge is generated in physics. 

Our proposed ontology of physics consists of three parts: system entities, magnitudes, and models.

\begin{figure}[htb]
\hspace{-2cm}
\begin{minipage}{\linewidth}
\begin{footnotesize}
\begin{tabular}{c}
\begin{forest}
forked edges,
for tree={draw,align=center}
[physics\\ entity
  [system\\ entity
   [system\\ element
    [\x{photon}]]
   [system\\ relation
    [\x{electromagnetic}\\ \x{interaction}]]
   [system
   [\x{Pfleegor-Mandel-}\\\x{photon}\\\x{system}]]]
   [magnitude
    [continuant\\ quality
     [\x{mass}]]
    [process\\ characteristic
     [\x{speed of}\\\x{light}]]]
   [model
    [textual\\ model]
    [graphic\\ model
      [\x{geometric}\\ \x{model}]]
    [mathematical\\ model 
     [\x{Pfleegor-Mandel-}\\\x{superposition}\\\x{equation}]]]
    ]
\end{forest}\\
\end{tabular}
\caption{Top-level entities of the ontology of physics; lowest nodes (\x{italics}) show only examples. The lines indicate the  genus-species-relation.}
\label{f:ontP}
\end{footnotesize}
\end{minipage}
\end{figure}
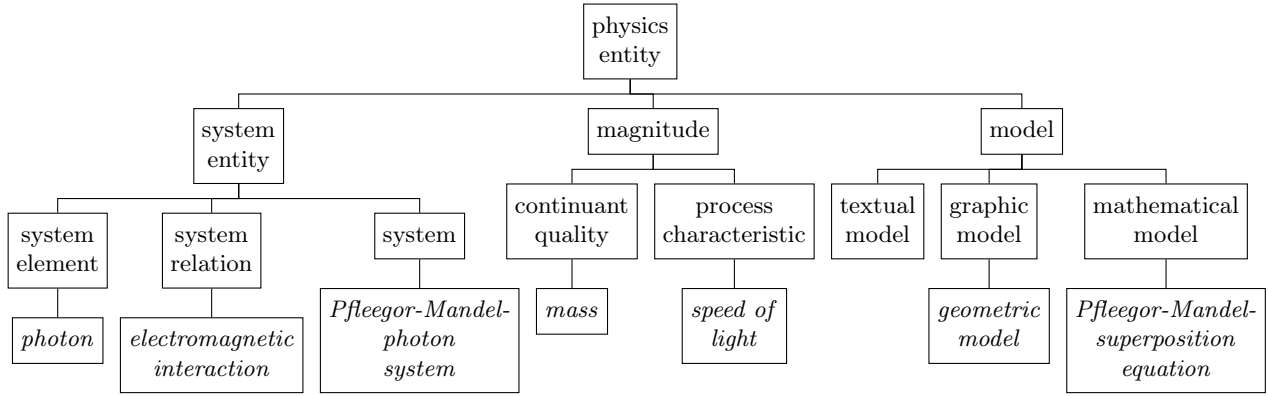

An overview is shown in figure \ref{f:ontP}, where we represent the ontology as a directed acyclic graph with nodes representing universals and idealities and edges representing binary relations between these entities, namely genus-species (class-subclass) relations. These function in the classical Aristotelian manner, namely a sub-entity $B$ of an entity $A$ has all the properties of $A$ plus at at least one property, called the specific difference. If the entities are idealities, they have no instances. For universals, however, the genus-species relation can be seen as equivalent to the instance-based relation according to which, for universals $A$ and $B$,  if $B$ is a subclass of $A$, every instance of $B$ is an instance of $A$.

\subsection{System entities} \label{sysE}
A \y{system} is a part of reality. It is a totality of entities (called `system elements' in figure \ref{f:ontP}), each made of matter, which interact with each other. To delimit a system is to select a range of granularity of elements and a system boundary \cite[ch.~8]{landgrebesmith:2025}.

Systems are always delimited by fiat, such as the solar system seen as a gravitational system. The sun itself can also be seen as a system of electromagnetic interactions, whose elements are the particles emitted by the sun.

In classical physics, all \y{system elements} are instances of Aristotelian universals. But when we zoom in using quantum physics, the system elements -- the particles -- lose their character as universals and become bound idealities. For example, though we can measure an individual photon, we cannot model it completely (see section \ref{einst}), rather we have only a wave equation that models the photon as the average of an ensemble. This is not only the case for the photon experiments described above (mirror equation (\ref{mirror}) and superposition equation (\ref{e:two})). It holds for all photons. For example, if a laser beam traverses two linear polarisers whose axes are turned against each other by an angle $\theta=\frac{\pi}{3},$ then the likelihood for each photon of passing the second polariser is $ \bra{x}\ket{x'} = \cos \theta~\mathrm{and~thus}~P_1=1-P_0=1-\cos^2 \theta=\frac{3}{4}$ so that 75\% of the laser's energy traverses the second polariser. But for a single photon, we cannot say whether it goes through \cite[Vol~3, ch.~11]{feynman:2010}. Thus also in this system, there is no individual photon as instance of a universal. The photon only corresponds in a certain sense (see section \ref{relP}) to the energy we measure, it is not a real entity but rather a bound ideality related to the measurement \x{and} to the mathematical free ideality which models the photon as a quantum wave.

The system elements of quantum physics are instanceless bound idealities, which are only connected to reality via attribution (see section \ref{declared}) and Bohr-correspondence. In Husserl's framework, to have evidence for the existence of entities of a given sort or for the truth of propositions of a given sort, we need to have a fulfilled intuition \x{(Anschauung)} that is provided by perception on the one hand or by another kind of intuition experienced by mathematicians when they grasp the validity of an argument \cite{smith:2000d}. Yet there is no \x{fulfilled mental act of perception or any other kind of intuition} of the photon or other system elements of quantum mechanics such as quarks or gluons.

The \y{system relations} of modern physics are the four interactions (electromagnetism, gravitation, the weak and the strong force). These are bound idealities because they are defined mathematically using the variants of QFT equations and gravitation theory (Einstein's field equations). The phenomena that we measure (the four forces) are then bound to those equations (in the sense of free ideality structures). But, and this shows to what extent they are bound idealities, the exact character of the bound idealities \x{weak} and \x{strong force} depend on the QFT variant we use. Thus, all system components of modern physics are bound idealities. In classical physics, which knows only gravitation and electromagnetism, these forces are of course universals.

\subsection{Magnitudes}

In physics, magnitudes are also called `physical quantities'. We, however, distinguish for clarity's sake between `magnitudes' on the one hand, and `quantities' on the other. A \textit{magnitude} is a phenomenon in reality -- for example mass, distance, velocity, acceleration or energy -- which has the feature that its instances can be quantified by \x{measurement}. Suppes' representational theory of measurement, \cite{suppes:1951, scottSuppes:1963,diez:1997} defines measurement as `the construction of mappings from empirical relational structures into numerical relational structures' \cite[p.~]{krantz:1971}\cite[sect.~3.4]{tal:2020}. Here, the empirical relational structure consists of a set of real objects (objects of measurement) which have qualitative relations among them.\footnote{The numerical relational structure consists of an algebraic field (such as $\mathbb{Z}$ or $\mathbb{Q}$) and constructors (such as $\leq$) or operators (such as addition) associated with them (an algebraic half-group). It is a scale. For example objects of different mass are mapped to a numerical scale, in this case with the constructors $>, <, \leq, \geq$ and the operator $+$. We follow this definition but say that the measurement yields a quantity.}

Magnitudes are always associated with material entities, which they quantify approximatively \cite[p.~46]{trout:1998}\cite[sect.~5]{tal:2020}. They are mind-independent properties of real entities \cite{byerly:1973, swoyer:1987, mundy:1987}. 

More precisely, magnitudes are either (i.) continuant magnitudes (such as mass or density), which are \x{qualities} of material bodies; or (ii.) process magnitudes, such as velocity or force, where the processes measured involve material bodies as participants and their interactions. Note that a continuant is an entity that persists, endures, or continues to exist through time while maintaining its identity \cite{bfo:2020}. Processes are occurrents, which means that they are entities that unfold through time, in other words entities that have temporal parts (parts along the temporal dimension). Each process magnitude at the instance level is what we call a process profile \cite{smith:2012c}. This means that it is a chain of process characteristics, for example a chain of values of the expression $\ddot{x}\vert_{t=j}$, for the acceleration of a mass determined at different time points $j$.

Magnitudes are assigned in each case to some particular. From the perspective of mathematical models in physics, we can also say that a magnitude is a dimension of a phase space, namely of the phase space which models the behaviour of the underlying particular in the modelled system \cite[sect.~7]{tal:2020}.

A physical \x{quantity}, now, in contrast to the magnitude, is the count of how many of this or that magnitude are found in a specific case of measurement, which we read from the measurement scale. It is that about which we speak when we collect measurement data and express it in terms of both units of measurement and quantities. Units of measurement are bound idealities since they are related to an amount of something real, but entirely conventional (culture-dependent), as can be seen immediately by examining some of the historical units of measurement for length, weight, or volume. 

\subsubsection{Attributed post-classical magnitudes}\label{declared}
The magnitude is something repeatable that we find in reality in many particulars. In classical physics, it is a universal, so that the property quantified by the magnitude exists both on the level of instances and on the level of universals and can be understood with the natural attitude of common sense.
Classical physics magnitudes, the associated processes of measurement, and the underlying qualities are all such as to fall within the realm of what can be imagined by our natural attitude towards the mesoscopic world which is the domain of common sense based thinking. 
In modern, post-classical physics, in contrast, for example in quantum physics, magnitudes no longer fall within the domain of what can be imagined using the natural attitude. As with system elements of quantum physics, so also with regard to these magnitudes we have no fulfilled acts of perception or non-mathematical intuition. Consider the quantum-theoretical magnitude \x{spin}. It is a \x{bound ideality} which relates a measurement to a mathematical \x{free ideality}, which is used to model the measurement in an equation.
For such magnitudes of modern physics, we have measurements (e.g. spin direction) for particular particles, but no corresponding universal. We thus have the following situation:

\begin{itemize}
 \item classical physics: $\langle \mathrm{universal,~instance} \rangle,$
 \item modern physics:  $\langle \mathrm{bound~ideality,~measurement~result} \rangle$,
\end{itemize}

where the instance of classical physics is replaced in modern physics by measurement result (modelled as a mathematical operator in quantum mechanics). The second type of tuple is a \x{matter of human declaration} (fiat attribution).
This is because, when we measure, for example, spin, there is nothing which we can identify as instance of a universal. In other words, the tuples of type $$\langle \mathrm{bound~ideality,~measurement~result} \rangle$$ are created by attribution because we are unable to understand the phenomena related to their associated magnitudes using the natural attitude, we cannot think of them as universals obtained by abstracting from their instances \cite[Inv.~6, \S 60]{husserl:2000}. Crucially, though we can measure individual particles, we cannot completely model them in a nonparadoxical way (see section \ref{einst}) since the wave function is either incomplete or forces us to accept action-at-a-distance. In other words, our models do not allow us to understand what it is that we measure, and thus we do not perform acts of abstraction to go from individuals to universals. Rather, in modern physics bound idealities take the place which universals have in classical physics. The instance of the magnitude is real, but it is represented by a measurement result (including the measurement error). Thus, magnitudes in modern physics can be understood only via mathematics, which means: through the use of models. Measurement results, however, are classical individuals (data) since they are produced by the machinery of modern physics. For example, the particle tracks in high energy physics which are determined using the classical Lorentz force, are instances of the universal \x{measurement result}, but this universal merely describes the technical nature of the observation, it does not link the individuals instantiating it to the models of physics.

Physicists \textit{model} the real magnitudes encountered in nature by means of mathematical structures which are models of natural processes. The magnitude itself, however, for example the spin, is an ideality bound through its relation to the measurement.

\x{Importantly,} each variable of the Schr{\"o}dinger equation (\ref{eeq}) has ontologically two aspects: it not not only represents a non-instantiable magnitude of \x{type bound ideality} (first aspect), but also a \x{mathematical, free ideality} (second aspect) in a human-created model (see section \ref{relP}). This second aspect of the expression becomes evident when physicists manipulate equations using mathematical operations  (as opposed to using the equation with measurement results to calculate a result). The equation itself, and the referring expressions within it, are free idealities, structures that are used by physicists to refer to measurable features of reality. These mathematical structures result from human intuition and are independent of any measurement process.

\subsection{Models}

A model in physics is a mind-dependent representation of an aspect of reality that uses abstract symbols and is created to describe, explain, or predict an aspect of reality. Abstract symbols are used primarily in the context of equations, but they may be accompanied by textual descriptions which may in turn be complemented by graphical, often geometrical, representations. 

In classical physics, we have systems in reality whose elements are instances of universals; we also have magnitudes, which have individual instances but are represented in our models as mathematical free idealities.
\x{In modern physics, in contrast, we encounter system entities, magnitudes and models which contain no universals, but only bound and free (mathematical) idealities}.  In sum, we cannot penetrate through the outer shell of mathematics to get at some `true nature' of what lies beyond our models. This is also the reason why there is such a plethora of arguments regarding quantum physics among physicists and philosophers.

But no matter whether we model in classical or modern physics, the models are always of mathematical nature and require an upper mathematics ontology to describe them. We have described this mathematics ontology in \cite{landgrebesmith:2023}.
Here, it suffices to say that all mathematical entities are free idealities. 

\subsection{Relations between the top level entities} \label{relP}

The relation between photon (a bound ideality), measurement result, magnitude and free ideality (mathematical model) is as follows: The photon ideality is bound to the measurement result (an instance of a universal, for example, an amount of detected energy) which has a magnitude (also a bound ideality). The magnitude is expressed mathematically by a symbol in a model equation, which is a free ideality. 

\section{Ontology based interpretations of quantum phenomena}

\subsection{Interpreting Einstein's dilemma}\label{dilemma}

The photon before its arrival at the mirror is modelled using the equation (\ref{mirror}):

$$\ket{\psi}=\frac{1}{\sqrt{2}}(\ket{\psi_1}+\ket{\psi_2}),$$
which we repeat here for convenience. As we have seen, this equation does not describe the reality which we measure, or, if we assume that it does, then this violates the locality principle. Einstein therefore believed that the equation is incomplete. In a way, he was right, but all attempts to improve the equation, to make it more complete -- such as David Bohm's\label{bohm} hidden variables -- failed.\footnote{Bohm's idea of modeling particles as infinitesimal points in his guide equation is not only incompatible with the Standard Model, which sees particles as having an extension, but also unable to model any magnitude other than position; it is also incompatible with relativity and QFT, and has recently been falsified experimentally \cite{sharoglazova:2025}.}

So in what sense was Einstein right? The Schr{\"o}dinger equation is a mathematical model of reality obtained using highly artificial technical settings. It was generated to predict properties of real quantum entities without understanding their behaviour. Therefore, it is necessarily incomplete. It is just a free ideality which allows us to calculate properties of these particles to which we have attributed bound idealities, the magnitudes of quantum physics. But the model (equation (\ref{e:two})) is \x{ontologically void}, which means that is cannot contribute to our understanding of the world from the perspective of the natural attitude, we cannot relate anything real to it for which we can get fulfillment of non-mathematical intentionality. In other words, the model does not help us to understand reality since it was created not based on understanding, but based on the mathematical modeling of relations between measurements.

\subsection{Interpreting the Pfleegor-Mandel-photon}\label{intSup}

The photon created by two lasers has the equation $$\Psi_1 \otimes \Psi_2 = \frac{1}{\sqrt{2}}(\ket{1_1}\ket{0_2} + \ket{0_1}\ket{1_2}).$$ Like the mirror photon equation of the previous example, it is ontologically void, we cannot learn anything about the reality of our world by looking at it. It describes neither a mereological character nor a causal path in spacetime for a particle. It does not represent the instance of a universal. \x{So what can we learn from this model?} That we have machines (like the setup of the Pfleegor-Mandel-experiment) which can create phenomena that we can measure and mathematically model, but not understand in the sense of fulfilment of an act of perception or non-mathematical intuition. 

In other words, though the Pfleegor-Mandel-photon is real, since we can measure it, it is not an instance of anything, but related to the \x{Pfleegor-Mandel superposition equation} shown in equation (\ref{e:two}) as explained in section \ref{relP}. The binding of the measured photon to the equation via its magnitude has the character of an attribution (sections \ref{declared} and \ref{relP}). Its energy, which allows us to detect it in the experiment, is real, and due to the detection method, we know it must be a photon. But the entity itself can only be described using references to the states $\ket{1_1}\ket{0_2}$ as superimposed with the states $\ket{0_1}\ket{1_2}$. Thus though the photon is real, all we can say about it is that it is bound by the ideality of its measurement magnitude (in this case, its energy) and described using the model equation (\ref{e:two}). 

\subsection{The measurement problem}

Since the interpretation of the measurement problem is often seen as a hallmark of an ontology of physics, we briefly describe how we stand in relation to this problem also. \textcite{busch:1991} see the measurement problem as the \x{objectification problem}, namely the `question of how definite measurement outcomes are obtained' (p.~1). They distinguish three main types of interpretations of quantum measurement: First there is (i) the statistical interpretation, for which the referents of quantum  measurement assertions are measurement outcomes (which does not lead to the objectification problem). And then there are two types of realistic interpretation – for which the referents are `object systems' (op. cit. p.~5) – which may be either  (ii) incomplete or (iii) complete. The objectification problem can in principle arise in relation to these two latter alternatives. However, it does not arise for (ii), because the latter excludes the objectification problem by asserting the incompleteness of the wave equation along the lines first postulated by Einstein \cite{einstein:1935}. It does, however, arise for (iii), but then the idea of complete object systems leads to the well-known paradoxes of quantum mechanics. 

Like Busch et al., we view quantum mechanics as being about measurement outcomes (option (i)). However, we deal with these outcomes in a different way, since we think that the quantum mechanical entities to which the measurement outcome relate are entities, but of a different kind than the entities to which the measurements of classical physics are related. We obtain knowledge -- such knowledge as we have -- of the quantum system entities in question by relating the measurements to bound idealities (magnitudes) and modeling them using free idealities (the equations we use as models representing what we measure, see equation \ref{mirror}). This binding is possible because the measurement outcomes harbour regularities which enable mathematical modelling. 
Measurement outcomes are bound to \x{entities given to us in mathematical intuitions}\footnote{This is just another kind of intuition, directed not at real entities, but at mathematical idealities.} via the chain of relations from the measurements to the free idealities of mathematics which we have described above. These idealities inform us about the way we think about and acquire true knowledge of the quantum phenomena we capture in measurement devices. But they do not inform us about how nature is constituted at the level of matter.\footnote{At the mesocopic level, as we saw in sectiobn \ref{declared}, we do obtain knowledge of the universals given in physics.}
Thanks to the introduction of idealities, our view is orthogonal to the three solution types distinguished by Busch and co-workers (and also to the three distinctions drawn by \textcite{maudlin:1995}). Introducing idealities shows how we can have an objective view in the sense of fulfilled mathematical intuition of the referents of quantum measurement results.

%% file: discussion.tex
\section{Discussion}\label{disc}

In this paper, we present an ontology of physics which divides reality into three types of entities: (i) Universals which have instances in reality, (ii) bound idealities, which are abstract and related to individual entities via convention or attribution, and (iii) free idealities, which do not exist in reality, and are abstract.
This ontology is necessary because, for the entities of modern physics, we do not obtain fulfilled acts of intuition (\x{Anschauung}) outside mathematics. Thus for example we cannot imagine spin without mathematics. Identifying these three types of entities allows us to avoid the following problems of many ontologies of physics:

\begin{enumerate}
 \item Confounding of ontology of nature with ontology of physics -- we clearly separate domains of nature selected by physicists for human enquiry (systems, section \ref{sysR}) from what we think about these domains instead of believing that mathematical models (free idealities) can replace our common sense understanding of reality, which is fundamentally wrong. 
 \item Involuntarily stumbling into the trap of an implicit idealism like \textcite{esfeld:2018} who interpret their model elements (which are infinitesimal points) as real (see section \ref{esf}).
 \item Viewing mind-dependent bound idealities as real, which is what occurs when post-classical magnitudes such as spin, colour charge, or units of measurement are seen as universals. For units of measurement, one can clearly see how absurd this view is, since our views of them have changed massively over the last hundred years alone and continue to change, proving that they are culture-dependent. For spin, the flaw is less obvious, but it becomes visible very clearly when looking at the Pfleefor-Mandel-photon (equation (\ref{e:two})).
\item Inability to give a convincing account of mathematical entities such as $\pi$, $e$, or the Pfleegor-Mandel superposition equation (\ref{e:two}). 
\end{enumerate}

By embracing Husserlian bound and free idealities in addition to universals, we obtain on the one hand an ontology of physics that is adequate for mind-independent natural entities such as a body, its qualities, its mass, and also to a real ray of light which we see by its illumination of snow flakes. All these entities are universals. But our ontology is adequate also for entities such as a single photon (a system element that is a bound ideality) or for an equation describing the likelihood of photons passing two polarisers set at an angle $\theta$ (a free ideality).

Note that we deviate from Husserl's view of both types of idealities in one respect. First, Husserl has two views of bound idealities: In his earlier phase, he called them categorical mixed essences (\x{kategorial vermischte Wesen}) \cite{pradelle:2020}, which are given not, as in the case of universals (through abstraction and variation), but rather through idealisation or predication \cite[I.,\S 18; VI., \S 41]{husserl:2000}. For example, to create a unit of measurement, we idealise an experienced entity, such as a gallon or a fathom, and use it to define a corresponding unit of volume or length, respectively.  Both are clearly bound idealities. Later Husserl called them `\x{gebundene Idealitäten}' \cite[\S~65]{husserl:1999}, which we translate here as `bound idealities'. They `convey a sense of the real and pertain to the real world [$\dots$]. Bound idealities are Earth-bound, Mars-bound, bound to certain territories' (loc. cit.). Clearly bound idealities have different relations to the real individuals they are related to depending on the subject matter. For example, the bound ideality `spin' is obtained from idealising experimental measurements and relating them (by using correspondence) to classical universals (in this case angular momentum). But the bound ideality `Post-Westphalian nation state' is obtained by relating an event (the Treaty of Westphalia in 1648) to another bound ideality (the nation state), which is bound to a real political order with certain properties.

As to free idealities: like Husserl, we see them as unchanging, omnitemporal, and acausal. He also saw them as mind-independent. While certainly the first three properties are indisputable, the mind-independence is problematic since it raises the question where these entities reside and how we access them. Husserl has no satisfying answer to this question. One possible view is that mathematical free idealities are \x{Denkmöglichkeiten} (possibilities of thinking) of the human mind which are obtained by deduction and intersubjective dialogue among mathematicians. For example, the ideality $\pi$ was discovered by thinking about the circumference and surface of the circle (another free ideality), and was then evolved further by Leibniz, who defined it infinitesimally, a concept further refined by Euler and Ramanujan.

The bound and free idealities of physics together allow us to understand how the human mind can utilise real experience to derive a mathematical body of knowledge. These idealities arise through the intuitions that we have when we analyse the results of experiments conducted in highly artificial and controlled settings. These enable us to create ever more sophisticated technical devices that enable ever more refined experiments. But idealities can be bound only to a tiny fraction of the world we live in \cite{cartwright:1983,cartwright:1999}; for most of reality they are missing. Even for a phenomenon such as a lightning storm  we do not have mathematical models and have to rely on our common sense. 

%% file: appendix.tex
\appendix \label{prop}
\section{Appendix: Ontologies of physics, an overview}
\subsection{Realist ontologies of physics}
There is in the current literature a broad spectrum of proposed ontologies of physics. We focus here on ontologies of matter, which can be classified into three broad groups: \x{realist, structuralist,} and \x{speculative}, for each of which we give examples.
\subsubsection{t'Hooft}
Perhaps the most radically \x{realist} approach to the ontology of matter is formulated by Gerard 't Hooft, who, in the tradition of Einstein, proposes a hidden variable model that postulates unobservable `fast variables' that would explain all the paradoxa of quantum mechanics \cite{thooft:2021}. In a manner that rests once more on the confounding of model and reality, he uses, the term `variables' to refer to real properties of matter which are at the same time mathematical entities. Here again, therefore, reality and model are confounded. 't Hooft believes that his fast variables deterministically cause the superposition we observe in quantum mechanics. He can thereby, he tells us,

\begin{quote} rescue the concept of ontology as opposed to epistemology in quantum mechanics [$\dots$] Atoms, molecules, electrons and other tiny entities are features of things that really exist. They evolve into different states or objects that also exist, according to universal physical laws.\cite{thooft:2021} \end{quote}

We see that t'Hooft claims that the idealities of modern physics `exist' in the same sense as macrosopic objects like tables or chairs exist. But this is not the case, as we saw in sections \ref{einst}f. He also claims that his theory overcomes non-locality (Bell's theorem). But unfortunately this theory is highly speculative and involves the troublesome thesis to the effect that there is a causal chain leading from the polarisation of a pair of entangled photons to the observer's choice of the measurement angles used to detect them \cite{falkenburg:2007}.

\subsubsection{Algebra-based ontologies}
The currently dominating so-called realist ontology of matter is based on the algebraic group structure, which is defined using symmetries in nature \cite{wigner:1939,newtonWigner:1949}. It is called `group structural realism' because it is the structure of the algebraic group that is seen as the foundation of the ontology \cite{roberts:2011}.\footnote{Group theory is also the foundation of the Standard Model since the latter classifies particles by their group-theoretical properties. Yet the SM is by itself ontologically neutral, since it does not make claims about the existence of the particles; it, too is a mathematical theory. This is not changed by the fact that it refers to particles.} In this algebraic approach, an elementary particle is regarded as an irreducible element of the group G of so-called `symmetries of nature', where the properties that display symmetries are not natural, but rather mathematically defined entities \cite{neemann:1991}\footnote{An algebraic group is a set of elements with an operation that maps to two elements a third element of the same set and fulfils the three group axioms (the operation is associative, has an identity element, and every element of the set has an inverse element). Note that `element here is used in the strict set-theoretical sense.}. So in this theory, a particle type is indirectly defined via mathematical models based on the experimental observations of the physicist. Thus it is not directly defined, in the way in which we define, for example, a type of tree using the properties we can observe in the world around us.

Another example of a structural realist particle-based ontology was recently proposed by \textcite{benitez:2023}, who regards particles as objects carrying interaction charges, where the objects are seen as bundles of properties. For example, an electron is seen as a bundle (in the sense of \cite{paul:2017}) of the properties mass, spin, and electroweak charge, a claim Benitez makes even in spite of the fact that the QFT framework, unlike quantum mechanics, cannot account for the existence of particles \cite[ch.~6.4.1]{falkenburg:2007}. The particles in this ontology are supposed to be real, and we obtain their characteristic properties by observing those of their interactions which reveal these properties. But since the particles are defined exclusively in terms of their mathematical properties, they cannot be real, but must be irreal. Or the theory is mistaken and then there are no such particles.

\subsection{A purely structuralist ontology of physics}\label{esf}
A recent example of \x{structuralism} is the minimalist structural ontology proposed by \textcite{esfeld:2018}, which postulates that the world is made of atoms. The authors view these as matter points, not to be understood in the sense of `atom' as in `hydrogen atom', but rather in the classical sense of Democritus. In their view, however, the atoms have no essence, and no essential properties. Instead, they are characterised (individuated) solely by the way in which they relate to other atoms. In this ontology, the sole relation that can be used to distinguish one atom from another is distance, which is the property that grounds the structural character of the ontology. The properties of atoms result solely from the dynamics of the changes in their mutual distance relations, so that `standing in the relations is all there is to theses objects -- the relations are their essence.' (p.~7). Of course, these point-like particles are theoretical entities, as the authors concede (p.~10). Such theoretical particles cannot qualify as entities of the natural world, since such entities have to be real. It seems that the proposed ontology is not a form of `super-Humeanism'\footnote{The authors think that they are more radical than David Hume, who, according to Lewis is `the great denier of necessary connections. [$\dots$] all there is to the world is a vast mosaic of locat matters of particular fact, just one little thing and then another.' \cite{lewis:1986}}, as the authors claim, but rather a form of involuntary neo-idealism since a genuinely realist ontology requires not theoretical but real entities. Yet the authors claim to be realists.

\subsection{Speculative ontologies of physics}
The \x{speculative} ontologies of matter (which 't Hooft calls `epistemology', by which he means `idealistic ontology') use the mathematical structures of QFT, the mathematical fields, as denizens of the real world. \textcite{baker:2009} gives an overview of such theories. Drawing an analogy to the measurement problem in quantum mechanics, he proposes that \begin{quote}a state in QFT represents propensities for the manifestation of certain (classical) fields in the event of measurement. Because each classical field configuration in a given superposition assigns field values to points, the expectation value of $\hat{\phi}(x)$ gives the mean expected value of the (classical) field strength at $x$.\footnote{Where $\hat{\phi}(x)$ is the operator-valued quantum field.} \cite[p.~590]{baker:2009}\end{quote}  What does this mean? The operator\footnote{An operator in quantum mechanics maps an element of a space of physical states onto another space of states.} is used to model a measurement. The principle is the same as in quantum mechanics, where, for example, a measurement is modelled via an operator when we utilise a Stern-Gerlach apparatus to measure a particle spin. In the same way, we also obtain a classical field by a measurement event in QFT. Here, the fields are seen as properties (magnitudes) of natural entities in the same way that spin is seen as an entity in quantum mechanics. \x{But neither are real entities, they are merely mathematical entities.} The basic problem of such theories is that they mistake a mathematical model for an entity that exists in reality, which means that they are not realist, but idealist ontologies, since mathematical models are made of mathematical entities, which are \x{free idealities} that do not exist in the real world.

%% file: hum.bib
@book{busch:1991,
  author={Busch, Paul and Lahti, Pekka J and Mittelstaedt, Peter},
  title={The Quantum Theory of Measurement},
  publisher={Springer},
  address = {Berlin},
  year={1991}
}

@article{sharoglazova:2025,
  title={Energy--speed relationship of quantum particles challenges Bohmian mechanics},
  author={Sharoglazova, Violetta and Puplauskis, Marius and Mattschas, Charlie and Toebes, Chris and Klaers, Jan},
  journal={Nature},
  volume={643},
  number={8070},
  pages={67--72},
  year={2025}
}

@article{huggett:2013,
  title={Emergent spacetime and empirical (in) coherence},
  author={Huggett, Nick and W{\"u}thrich, Christian},
  journal={Studies in History and Philosophy of Science Part B: Studies in History and Philosophy of Modern Physics},
  volume={44},
  number={3},
  pages={276--285},
  year={2013}
}

@incollection{simons:1992,
  title={The formalization of Husserl’s theory of wholes and parts},
  author={Simons, Peter},
  booktitle={Philosophy and Logic in Central Europe from Bolzano to Tarski: Selected Essays},
  pages={71--116},
  year={1992},
  publisher={Springer}
}

@article{smith:1986,
  title={Husserl’s \x{Logical Investigations}},
  author={Smith, Barry and Mulligan, Kevin},
  journal={Grazer Philosophische Studien},
  volume={27},
  number={1},
  pages={199--207},
  year={1986},
  publisher={Brill}
}

@article{pradelle:2020,
  title={Anschauung und Idealit{\"a}ten},
  author={Pradelle, Dominique},
  journal={Ph{\"a}nomenologische Forschungen},
  number={1},
  pages={137--166},
  year={2020}
}

@article{einstein:1935,
  title={Can quantum-mechanical description of physical reality be considered complete?},
  author={Einstein, Albert and Podolsky, Boris and Rosen, Nathan},
  journal={Physical review},
  volume={47},
  number={10},
  pages={777},
  year={1935}
}

@article{maudlin:1995,
  title={Three measurement problems},
  author={Maudlin, Tim},
  journal={Topoi},
  volume={14},
  number={1},
  pages={7--15},
  year={1995}
}

@book{sider:2001,
  title={Four-dimensionalism},
  author={Sider, Theodore},
  year={2001},
  publisher={Oxford University Press}
}

@incollection{simons:2002,
  title={Candidate General Ontologies for Situating Quantum Field Theory},
  author = {Peter Simons},
  booktitle={Ontological aspects of quantum field theory},
  editor={Kuhlmann, Meinard and Lyre, Holger and Wayne, Andrew},
  year={2002},
  pages={33--55},
  publisher={World Scientific}
}

@incollection{kuhlmann:2002,
  booktitle={Ontological aspects of quantum field theory},
  editor={Kuhlmann, Meinard and Lyre, Holger and Wayne, Andrew},
  year={2002},
  publisher={World Scientific}
}

@book{falkenburg:2007,
  title={Particle metaphysics: A critical account of subatomic reality},
  author={Falkenburg, Brigitte},
  year={2007},
  address={Berlin},
  publisher={Springer}
}

@article{baker:2009,
  title={Against field interpretations of quantum field theory},
  author={Baker, David John},
  journal={The British Journal for the Philosophy of Science},
  volume ={60},
  number={3},
  pages = {585-609},
  year={2009},
  publisher={The University of Chicago Press}
}

@article{benitez:2023,
  title={Particles, fields, and the ontology of the standard model},
  author={Benitez, Federico},
  journal={Synthese},
  volume={201},
  number={1},
  pages={20},
  year={2023}
}

@article{thooft:2021,
  title={Ontology in quantum mechanics},
  author={Hooft, Gerard't},
  journal={Topics on Quantum Information Science},
  volume={5},
  pages={13},
  year={2021}
}

@incollection{paul:2017,
  booktitle= {Being, Freedom, and Method: Themes from the Philosophy of Peter van Inwagen},
  title={A one category ontology},
  author={Paul, Laurie A},
  editor ={Keller, John A},
  publisher = {Oxford Scholarship Online},
  year={2017}
}

@book{esfeld:2018,
  title={A minimalist ontology of the natural world},
  author={Esfeld, Michael and Deckert, Dirk-Andr{\'e}},
  year={2017},
  publisher={Routledge}
}

@article{landgrebesmith:2023,
      title={Ontologies of common sense, physics and mathematics},
      author={Jobst Landgrebe and Barry Smith},
      year={2023},
      journal={arXiv:2305.01560}
}

@Book{		  cartwright:1999,
  title		= {The dappled world: A study of the boundaries of science},
  author	= {Cartwright, Nancy},
  year		= {1999},
  publisher	= {Cambridge University Press}
}

@article{roberts:2011,
  title={Group structural realism},
  author={Roberts, Bryan W},
  journal={The British Journal for the Philosophy of Science},
  volume = {62},
  pages={47-69},
  year={2011}
}

@article{neemann:1991,
  title={Internal supersymmetry and superconnections},
  author={Ne'eman, Yuval and Sternberg, Shlomo and Donato, P and Duval, C and Elhadad, J and Tuynman, G},
  journal={Symplectic Geometry and Mathematical Physics},
  pages={326--54},
  year={1991}
}

@article{wigner:1939,
  title={On unitary representations of the inhomogeneous {L}orentz group},
  author={Wigner, Eugene},
  journal={Annals of mathematics},
  pages={149--204},
  year={1939}
}

@article{scottSuppes:1963,
  title={Foundational Aspects of Theories of Measurement},
  author={Scott, Dana and Suppes, Patrick},
  journal={Readings in Mathematical Psychology.(Edited by R. Duncan Luce, Robert R. Bush, and Eugene Galanter.) New York: John Wiley \& Sons},
  volume={1},
  pages={212--27},
  year={1963}
}

@article{suppes:1951,
  title={A set of independent axioms for extensive quantities},
  author={Suppes, Patrick},
  journal={Portugaliae Mathematica},
  volume={10(4)},
  pages= {163-172},
  year={1951}
}

@article{diez:1997,
  title={A hundred years of numbers. an historical introduction to measurement theory 1887--1990: Part I: the formation period. two lines of research: axiomatics and real morphisms, scales and invariance},
  author={D{\'i}ez, Jos{\'e}A},
  journal={Studies in History and Philosophy of Science Part A},
  volume={28},
  number={1},
  pages={167--185},
  year={1997}
}

@book{krantz:1971,
  title={Foundations of measurement, Vol. I: Additive and polynomial representations},
  author={Krantz, David and Luce, Duncan and Suppes, Patrick and Tversky, Amos},
  publisher={Academic Press},
  year={1971}
}

@article{newtonWigner:1949,
  title={Localized states for elementary systems},
  author={Newton, Theodore Duddell and Wigner, Eugene P},
  journal={Reviews of Modern Physics},
  volume={21},
  number={3},
  pages={400},
  year={1949}
}

@Book{		  cartwright:1983,
  title		= {How the Laws of Physics Lie},
  author	= {Nancy Cartwright},
  year		= {1983},
  publisher	= {Oxford University Press},
  address	= {Oxford}
}

@Article{	  smith:2022,
  title		= {The birth of ontology},
  author	= {Smith, Barry},
  journal	= {Journal of Knowledge Structures and Systems},
  volume	= {3},
  number	= {1},
  pages = {57-66},
  year		= {2022}
}

@Book{		  bfo:2020,
  title		= {Information technology - Top-level ontologies (TLO) - Part 2: Basic Formal Ontology (BFO)},
  author	= {ISO/IEC 21838-2.2},
  publisher	= {International Standardization Organization},
  address	= {New York},
  year		= {2020}
}

@article{norsen:2005,
  title={Einstein’s boxes},
  author={Norsen, Travis},
  journal={American Journal of Physics},
  volume={73},
  number={2},
  pages={164--176},
  year={2005}
}

@article{ballentine:1972,
  title={Einstein’s interpretation of quantum mechanics},
  author={Ballentine, Lelslie E},
  journal={American Journal of Physics},
  volume={40},
  number={12},
  pages={1763--1771},
  year={1972}
}

@book{  jammer:1974,
        title = {The philosophy of Quantum Mechanics: The interpretations of Quantum Mechanics in historical perspective.},
        author =  {Jammer, Max},
        publisher = {John Wiley and Sons},
        address= {New York},
        year =  {1974}
}

@Book{		  feynman:2010,
  title		= {The {F}eynman Lectures on Physics (1964)},
  author	= {Richard P. Feynman and Robert B. Leighton and Matthew
		  Sands},
  publisher	= {Addison-Wesley},
  address	= {Boston, MA},
  year		= {2010}
}

@InCollection{tal:2020,
	author       =	{Tal, Eran},
	title        =	{{Measurement in Science}},
	booktitle    =	{The {Stanford} Encyclopedia of Philosophy},
	editor       =	{Edward N. Zalta},
	howpublished =	{\url{https://plato.stanford.edu/archives/fall2020/entries/measurement-science/}},
	year         =	{2020}
}

@article{mundy:1987,
  title={The metaphysics of quantity},
  author={Mundy, Brent},
  journal={Philosophical Studies},
  volume={51},
  pages={29--54},
  year={1987}
}

@incollection{swoyer:1987,
  title={The metaphysics of measurement},
  author={Swoyer, Chris},
  booktitle={Measurement, realism and objectivity: Essays on measurement in the social and physical sciences},
  pages={235--290},
  year={1987},
  publisher={Springer}
}

@article{byerly:1973,
  title={Realist foundations of measurement},
  author={Byerly, Henry C and Lazara, Vincent A},
  journal={Philosophy of Science},
  volume={40},
  number={1},
  pages={10--27},
  year={1973}
}

@book{trout:1998,
  title={Measuring the intentional world: Realism, naturalism, and quantitative methods in the behavioral sciences},
  author={Trout, John D},
  year={1998},
  publisher={Oxford University Press}
}

@Article{	  smith:2010,
  title		= {Ontological Realism: A Methodology for Coordinated
		  Evolution of Scientific Ontologies},
  author	= {Smith, Barry and Ceusters, Werner},
  journal	= {Applied Ontology},
  volume	= {5},
  number	= {3-4},
  pages		= {139--188},
  year		= {2010},
  publisher	= {IOS Press}
}

@book{lewis:1986,
  title={Philosophical Papers, Volume 2: Mind, Language, Epistemology},
  author={Lewis, David},
  year={1986},
  publisher={Oxford University Press}
}

@Book{		  scheler:1973,
  title		= {Formalism in Ethics and Non-Formal Ethics of Values},
  year		= {1973 [1913/1916]},
  publisher	= {Northwestern University Press},
  address	= {Evanston},
  author	= {Max Scheler}
}

@Book{		  ingarden:1973,
  title		= {The Literary Work of Art. {I}nvestigations on the
		  Borderlines of Ontology, Logic and the Theory of
		  Literature},
  author	= {Ingarden, Roman},
  publisher	= {Northwestern University Press},
  address	= {Evanston, IL},
  year		= {1973}
}

@Book{		  landgrebesmith:2025,
  title		= {Why machines will never rule the world. AI without fear.},
  author	= {Landgrebe, Jobst and Smith, Barry},
  publisher	= {Routledge},
  edition = {2nd ed.},
  address	= {London},
  year		= {2025}
}

@book{erhard:2014,
  title={Denken {\"u}ber nichts-Intentionalit{\"a}t und Nicht-Existenz bei Husserl},
  author={Erhard, Christopher},
  year={2014},
  publisher={Walter de Gruyter GmbH \& Co KG}
}

@book{husserl:1975,
  title={Experience and judgment},
  author={Husserl, Edmund},
  year={1975},
  publisher={Northwestern University Press}
}

@book{  husserl:1988,
        title = {Vorlesungen Über Ethik und Wertlehre 1908–1914. Husserliana XXVIII},
        author =  {Husserl, Edmund},
        editor={Ullrich Melle},
        publisher = {Kluwer},
        address= {Dordrecht},
        year =  {1988}
}

@book{husserl:1966,
  title={Analysen zur passiven Synthesis: aus Vorlesungs-und Forschungsmanuskripten 1918-1926},
  author={Husserl, Edmund and Fleischer, Margot and van Breda, Herman L},
  volume={XI},
  year={1966},
  publisher={Kluwer},
  address= {Dordrecht}
}

@book{  husserl:1999,
        title = {Erfahrung und Urteil},
        author =  {Husserl, Edmund},
				editor = {Landgrebe, Ludwig},
        publisher = {Felix Meiner},
        address= {Hamburg},
        year =  {1999 [1938]}
}

@Misc{		  feynman:1960,
  author	= {Feynman, Richard},
  title		= {Knowing versus understanding},
  year		= {1960},
  url		= {https://www.youtube.com/watch?v=NM-zWTU7X-k}
}

@Book{		  husserl:2000,
  title		= {Logical {I}nvestigations},
  year		= {2000 [1901]},
  author	= {Edmund Husserl},
  publisher	= {Routledge},
  address	= {Abingdon}
}

@Article{	  smith:2012c,
  title		= {Classifying processes: an essay in applied ontology},
  author	= {Smith, Barry},
  journal	= {Ratio},
  volume	= {25},
  number	= {4},
  pages		= {463--488},
  year		= {2012}
}
